\documentclass[12pt]{article}%
\usepackage{amsfonts}
\usepackage{graphicx}
\usepackage{amsmath}
\usepackage{amssymb}%
\pdfoutput=1
\makeatletter

\@addtoreset{equation}{section}
\makeatother
\newcommand{\beq}{\begin{equation}}
\newcommand{\eeq}{\end{equation}}
\renewcommand{\a}{\alpha}

\begin{document}

%\tableofcontents

\baselineskip=18pt
%a la harvmac
\baselineskip 0.7cm

\begin{titlepage}
%% Set the number of the title with 0
\setcounter{page}{0}
% change the footnote symbol
\renewcommand{\thefootnote}{\fnsymbol{footnote}}
%------------------
\begin{flushright}
%preprint number
%CALT-mm-nnnn\\
%IPMU09-nnnn\\
%UT-09-30\\
%month, 2008
\end{flushright}
%---------------------
\vskip 1.5cm
\begin{center}
{\LARGE \bf
Integral Representations on Supermanifolds: super Hodge duals, PCOs and Liouville forms.
\vskip 1.5cm
}
{
Leonardo Castellani$^{~a,b,}$\footnote{leonardo.castellani@uniupo.it},
Roberto Catenacci$^{~a,c,}$\footnote{roberto.catenacci@uniupo.it},
and
Pietro Antonio Grassi$^{~a,b,}$\footnote{pietro.grassi@uniupo.it}
\medskip
}
\vskip 0.5cm
{
\small\it
\centerline{$^{(a)}$ Dipartimento di Scienze e Innovazione Tecnologica, Universit\`a del Piemonte Orientale} }
\centerline{\it Viale T. Michel, 11, 15121 Alessandria, Italy}
\medskip
\centerline{$^{(b)}$ {\it
INFN, Sezione di Torino, via P. Giuria 1, 10125 Torino} }
\centerline{$^{(c)}$ {\it
Gruppo Nazionale di Fisica Matematica, INdAM, P.le Aldo Moro 5, 00185 Roma} }
\vskip  .5cm
\medskip
\end{center}
%-----------------------------------------
\centerline{{\bf Abstract}}
\medskip
\noindent
We present a few types of integral transforms and integral representations that are very useful for extending to supergeometry many familiar concepts of differential geometry. Among them we discuss the construction of the super Hodge dual, the integral representation of picture changing operators of string theories and the construction of the super-Liouville form of a symplectic supermanifold. \\

\bigskip
\bigskip
\noindent
\bigskip

\medskip
%\today
\end{titlepage}\setcounter{page}{1}
%don't number title page

\vfill
\eject

\newpage\setcounter{footnote}{0} \newpage\setcounter{footnote}{0}
%%%%%%%%%%%%%%%%%%%%%%%%%%%%%%%%%%%%%%%%%%%%%%%%%%%%%%%%%%%%%%%%%%%%%%%

\section{Introduction}

Pursuing the construction of supersymmetric Lagrangians based in the framework
of supermanifold geometry, we proposed in \cite{Castellani:2015paa} a new
\textit{Hodge operator} $\star$ acting on (super)differential forms. For that
aim, we have discussed a complete formalism (integral-, pseudo- e super-forms,
their complexes and the integration theory) in a series of papers
\cite{Castellani:2015paa,Castellani:2014goa,Castellani:2015ata} together with
a suitable Hodge operator.

As a byproduct, this mathematical tool sheds also a new light on the Hodge
operator in conventional differential geometry.

The theory of forms in supergeometry was extensively developed from a
mathematical point of view (important general references are \cite{Manin},
\cite{Deligne}, \cite{Rogers}, \cite{Bruzzo}); the main aim of this paper is
to present, in a formalism that make contact with the recent physical
literature, a few subjects of interest mainly in physical applications.

The Hodge operator plays an essential role in differential geometry, yielding
a fundamental relation between the exterior bundle of differential forms and
the scalar product $(\bullet,\bullet)$ on the manifold. The construction
requires the existence of a metric $g$ on the the manifold $\mathcal{M}$ and
is an involutive operation $\star$ which satisfies the linearity condition
$\star(f\omega)=f\star\omega$ with $\omega$ a given $p$-form.

In the case of supermanifolds (we refer for ex. to \cite{Castellani:2015paa}
for the basic ingredients of supergeometry with the notations and the
definitions used in this paper; see also \cite{Witten:2012bg} for a recent
extensive review), the definition of the Hodge dual turns out to be harder
than expected since one has to deal with the infinite-dimensional complexes of
superforms. The integral forms and pseudo-forms are crucial to establish the
correct matching of elements between the different spaces of forms. This new
type of differential forms requires the enlargement of the conventional space
spanned by the fundamental $1$-forms, admitting distribution-like expressions
(essentially, Dirac delta functions and Heaviside step functions). This has
triggered us to consider the \textit{Fourier analysis} for differential forms
(this was earlier considered in \cite{VORONOV1} and \cite{Kalkman:1993zp}),
and leads to an integral representation of the Hodge operator as explained in
\cite{Castellani:2015paa,Castellani:2015ata}. Such a representation can first
be established in the case of a conventional manifold $M$ without any
reference to supermanifolds, except for the notion of \textit{Berezin
integral}. A new set of anticommuting variables playing the role of dual
variables to fundamental 1-forms $dx^{i}$ is introduced and the Hodge operator
is defined by a suitable Berezin integration on the new variables. The result
is proven to coincide with the usual Hodge operator. When extended to
supermanifolds, our construction yields a \textquotedblleft good" definition
of Hodge operator, satisfying all desired properties. Note that our fiberwise
integral transform maps superforms into integral forms showing that the latter
are fundamental ingredients for the Hodge theory in supermanifolds.

In the case of supermanifolds Voronov and Zorich studied in \cite{VORONOV1} a
different type of fiberwise integral transform, that in the purely bosonic
setting (usual differential forms) also gives the usual Hodge dual. Their
transform maps usual differential forms defined on the parity reversed (i.e.
even) normal bundle of the reduced bosonic manifold (embedded into the
supermanifold setting to zero the anticommuting coordinates) to pseudoforms on
the odd normal bundle. Depending on the nature of the forms transformed, in
this setting the pseudoforms obtained are very general: polynomial and
analytic forms in the commuting differentials and even distribution-like forms.

This type of transform is not suitable for extending the Hodge dual to
supergeometry but it is very interesting because, in modern string theory
language, it is strictly related to (the integral representation of) the
picture changing operators.

As is well known, in conventional Fourier analysis, the Fourier transform of a
product of two functions is the \textit{convolution integral} of the Fourier
transform of the two functions. In a reciprocal way, the Fourier transform of
the convolution integral of two functions can be expressed as the product of
the Fourier transforms of the latter. This simple formula can be imported in
our framework where the Fourier transform represents the Hodge operator and
the convolution integral is a suitable Berezin integral of two differential
forms. With this observation we are able to express the Hodge dual of the
wedge product of two differential forms as the (Berezin)-convolution of the
Hodge duals of the differential forms.

In the case of integral forms (that are "distribution-like") the convolution
product is clearly the right one, because their graded wedge product vanishes.

Another problem encountered in extending to supermanifolds the concepts of the
usual differential geometry is that of the Liouville form in a symplectic
supermanifold. The problem here is that the super-symplectic form is naturally
a superform of zero picture and hence the Liouville form cannot be defined
simply as a graded exterior power of the symplectic form, because there is no
a top exterior power and the correct Liouville form must be defined using
instead integral forms. Also in this case an integral representation opens the way.

In this present paper we will elaborate on these subjects, on their
definitions and properties.

Finally, the Fourier (Berezin) integral representation of the Hodge dual
operator can be extended to noncommutative spaces. A very recent work on this
appeared in \cite{Majid}, and the idea of a Fourier-Berezin transform can be
found also in \cite{Majid2}.

\section{Forms and Integration}

The usual integration theory of differential forms for bosonic manifolds can
be conveniently rephrased to uncover its relation with Berezin integration
\cite{Bernstein-Leites-How},\cite{Voronov-geometric}.

We start with a simple example: consider in $\mathbb{R}$ the integrable 1-form
$\omega$ $=g(x)dx$ (with $g(x)$ an integrable function in $\mathbb{R}$ ). We
have:
\[
\int_{\mathbb{R}}\omega=\int_{\mathbb{-\infty}}^{+\infty}g(x)dx\,.
\]
Observing that $dx$ is an anticommuting quantity, and denoting it by $\psi$,
we could think of $\omega$ as a function on the superspace $\mathbb{R}^{1|1}$:%
\begin{equation}
\omega=g(x)dx=f(x,\psi)=g(x)\psi
\end{equation}
This function can be integrated \textit{\`{a} la} Berezin reproducing the
usual definition:%
\[
\int_{\mathbb{R}^{1|1}}f(x,\psi)[dxd\psi]=\int_{\mathbb{-\infty}}^{+\infty
}g(x)dx\,=\int_{\mathbb{R}}\omega
\]
Note that the symbol of the formal measure $[dxd\psi]$ is written just to
emphasize that we are integrating on the \textbf{two} variables $x$ and $\psi
$, hence the $dx$ inside $[dxd\psi]$ is \textit{not} identified with $\psi$.

Denoting by $M$ a bosonic orientable differentiable manifold of dimension $n$,
its exterior bundle $\bigwedge(M)=\sum_{p=0}^{n}\bigwedge^{p}(M)$ is the
direct sum of $\bigwedge^{p}(M)$ (their spaces of sections will be denoted as
$\Omega(M)$ and $\Omega^{p}(M)$ respectively). A section $\omega$ of
$\bigwedge^{p}(M)$ can be written locally as
\begin{equation}
\omega=\sum\omega_{i_{1}\dots i_{p}}(x)dx^{i_{1}}\wedge\dots\wedge dx^{i_{p}}
\label{inA}%
\end{equation}
where the coefficients $\omega_{i_{1}\dots i_{p}}(x)$ are functions on $M$ and
$i_{1}<...<i_{p}$. The integral of $\omega\in\Omega^{n}(M)$ is defined as:
\begin{equation}
I[\omega]=\int_{M}\omega=\int_{M}\omega_{12...n}(x)\,d^{n}x\,, \label{inB}%
\end{equation}
This opens the way to relating the integration theory of forms and the Berezin
integral, by substituting every $1$-form $dx^{i}$ with a corresponding
abstract Grassmann variable denoted again with $dx^{i}.$ A section $\omega$ of
${\Omega(}M{)}$ is viewed locally as a function on a supermanifold
$\mathcal{M=}T^{\ast}(M)$ with coordinates $(x^{i},dx^{i}):$
\begin{equation}
\omega(x,dx)=\sum\omega_{i_{1}\dots i_{p}}(x)dx^{i_{1}}\dots dx^{i_{p}}\,;
\label{inAA}%
\end{equation}
such functions are polynomials in $dx$'s. Supposing now that the form $\omega$
is integrable, its Berezin integral gives:
\begin{equation}
\int_{\mathcal{M=}T^{\ast}(M)}\omega(x,dx)[d^{n}xd^{n}(dx)]=\int_{M}\omega
\end{equation}

\section{The Integral Representation of the Hodge Star and Convolutions}

In the following, for a given set $\{\xi^{i}\}_{i=1}^{n}$ of Grassmann
variables, our definition of the Berezin integral is $\int_{\mathbb{R}^{0|n}%
}\xi^{1}...\xi^{n}\left[  d^{n}\xi\right]  =1$ and not $\int_{\mathbb{R}%
^{0|n}}\xi^{1}...\xi^{n}\left[  d^{n}\xi\right]  =\left(  -1\right)
^{\frac{n(n-1)}{2}}.$ Moreover, if $\alpha$ is a monomial expression of some
anticommuting variables $\alpha^{k}$ not depending on the $\xi^{i},$ we
define: $\int_{\mathbb{R}^{0|n}}\alpha\xi^{1}...\xi^{n}\left[  d^{n}%
\xi\right]  =\alpha,$ where the product between $\alpha$ and the $\xi^{i}$ is
the usual $\mathbb{Z}_{2}$ graded wedge product in the superalgebra generated
by the tensor product of the Grassmann algebra generated by the $\xi^{i}$ and
that generated by the $\alpha^{k}:$ if $\mathcal{A}$ and $\mathcal{B}$ are two
$\mathbb{Z}_{2}$-graded algebras with products $\cdot_{\mathcal{A}}$and
$\cdot_{\mathcal{B}}$, the $\mathbb{Z}_{2}$-graded tensor product
$\mathcal{A}\otimes\mathcal{B}$ is a $\mathbb{Z}_{2}$-graded algebra with the
product (for homogeneous elements) given by :%
\[
(a\otimes b)\cdot_{\mathcal{A}\otimes\mathcal{B}}(a^{\prime}\otimes b^{\prime
})=(-1)^{\left\vert a^{\prime}\right\vert \left\vert b\right\vert }%
a\cdot_{\mathcal{A}}a^{\prime}\otimes b\cdot_{\mathcal{B}}b^{\prime}%
\]
In our case the algebras are Grassmann algebras and the products $\cdot$ are
wedge products. The symbols $\otimes$ and $\wedge$ will be, in general, omitted.

One can observe, see {\it e.g.} \cite{Berezin79}, \cite{VORONOV99}, and also
\cite{Castellani:2015paa} that the usual Hodge dual in $\mathbb{R}^{n}%
$\textbf{ }(for a metric\footnote{We consider metrics of arbitrary signature;
The sign of $g$ fixes the sign of the overall coefficient $\frac
{\sqrt{\left\vert g\right\vert }}{g}=\pm\frac{1}{\sqrt{\left\vert g\right\vert
}}.$} given by a matrix $A$ with entries $g_{ij}$ ) can be obtained by means
of\textbf{ }the Fourier (Berezin)-integral transform $\mathcal{T}$ . For
$\omega(x,dx)\in\Omega^{k}(\mathbb{R}^{n})$ we have:
\begin{equation}
\star\omega=i^{\left(  k^{2}-n^{2}\right)  }\frac{\sqrt{\left\vert
g\right\vert }}{g}\mathcal{T}(\omega)=i^{\left(  k^{2}-n^{2}\right)  }%
\frac{\sqrt{\left\vert g\right\vert }}{g}\int_{\mathbb{R}^{0|n}}\omega
(x,\eta^{\prime})e^{idxA\eta^{\prime}}[d^{n}\eta^{\prime}] \label{duale1}%
\end{equation}
where $g=\mathrm{det}A.$ The exponential series defining $e^{idxA\eta^{\prime
}}$ is written using the $\mathbb{Z}_{2}$ graded wedge product quoted above.
The Grassmann variables $\eta^{\prime}$ are defined as $\eta^{\prime}%
=A^{-1}\eta$ where the $\eta$ are the \textit{(parity changed)}\footnote{Here
and in the following we adopt the convention that $d$ is an odd operator (so
$dx$ is a odd form but $\frac{\partial}{\partial x}$ is an even vector). A
change of parity is necessary because we want $\eta$ to be an odd variable.}
variables dual to the $dx$. In this way the covariance properties of
$\omega(x,\eta^{\prime})$ are exactly those of a differential form and this is
crucial in physical applications.

The factor $i^{\left(  k^{2}-n^{2}\right)  }$ can be obtained by computing the
transformation of the monomial form $dx^{1}dx^{2}...dx^{k}\ $in the simple
case $A=I.$

The explicit computation gives:%
\begin{equation}
i^{\left(  k^{2}-n^{2}\right)  }\mathcal{T}\left(  dx^{1}...dx^{k}\right)
=\star\left(  dx^{1}...dx^{k}\right)
\end{equation}
and%
\begin{equation}
\mathcal{T}^{2}\left(  \omega\right)  =i^{\left(  n^{2}-k^{2}\right)
}i^{\left(  k^{2}\right)  }\left(  \omega\right)  =i^{n^{2}}\left(
\omega\right)  \label{quadratodifourier}%
\end{equation}
yielding the usual duality relation:%
\begin{equation}
\star\star\omega=i^{(\left(  n-k)^{2}-n^{2}\right)  }i^{\left(  k^{2}%
-n^{2}\right)  }i^{n^{2}}(\omega)=(-1)^{k(k-n)}(\omega)
\end{equation}

As for functions, one can define (see \cite{VORONOV1}) a convolution product
between differential forms on an ordinary manifold. The starting point is
again the interpretation of differential forms as functions of the commuting
variables $x$ and the anticommuting variables $dx.$ For $\alpha\in\Omega
^{p}(\mathbb{R}^{n})$ and $\beta\in\Omega^{q}(\mathbb{R}^{n}),$ the
convolution product $\bullet$ is defined using Berezin integration on the
anticommuting variables:%
\begin{equation}
\alpha\bullet\beta(x,dx)=\int_{\mathbb{R}^{0|n}}\alpha(x,\xi)\beta
(x,dx-\xi)[d^{n}\xi] \label{convoluzione}%
\end{equation}
where the $\xi$ are auxiliary anticommuting variables. Note that this
pointwise convolution product depends on a choice of a volume element (i.e.
the ordering of the auxiliary variables). The convolution product \footnote{We
must integrate generically monomials of the type $\left(  \xi\right)
^{p+q-k}\left(  dx\right)  ^{k}$ and the Berezin integration selects
$k=p+q-n.$} maps $\Omega^{p}\times\Omega^{q}\rightarrow\Omega^{p+q-n}$ . To
obtain (generically) non trivial results we must have $0\leq p+q-n\leq n.$ The
algebra of this convolution is
\[
\alpha\bullet\beta=(-1)^{\left(  n^{2}+pq\right)  }\beta\bullet\alpha
\]
The convolution `interacts" well with the integral transformation
$\mathcal{T}$ defined above and the wedge product. We will consider explicitly
only the standard bosonic case in which the matrix $A$ of the.previous
paragraph is the identity matrix $I.$

For example, in the case $n=4,$ we can compute $\mathcal{T}\left(
dx^{1}dx^{2}\right)  =dx^{3}dx^{4}$ and $\mathcal{T}\left(  dx^{1}\right)
=\left(  -i\right)  dx^{2}dx^{3}dx^{4},$ $\mathcal{T}\left(  dx^{2}\right)
=idx^{1}dx^{3}dx^{4}.$ The convolution is:%
\[
\mathcal{T}\left(  dx^{1}\right)  \bullet\mathcal{T}\left(  dx^{2}\right)
=\int_{\mathbb{R}^{0|4}}\left(  -i\right)  \xi^{2}\xi^{3}\xi^{4}\left(
i\right)  \left(  dx^{1}-\xi^{1}\right)  \left(  dx^{3}-\xi^{3}\right)
\left(  dx^{4}-\xi^{4}\right)  [d^{4}\xi]
\]
\[
=dx^{3}dx^{4}=\mathcal{T}\left(  dx^{1}dx^{2}\right)
\]
Another simple example is the case $q=n-p$ where we find:%
\begin{equation}
i^{n^{2}}\left(  -1\right)  ^{p}\left(  -1\right)  ^{p\left(  n-p\right)
}\mathcal{T}\left(  \alpha\beta\right)  =\mathcal{T}\left(  \alpha\right)
\bullet\mathcal{T}\left(  \beta\right)  \label{convoluzione1}%
\end{equation}

\noindent Indeed, recalling that:
\begin{equation}
\mathcal{T}\left(  dx^{1}...dx^{p}\right)  =i^{\left(  n^{2}-p^{2}\right)
}(dx^{p+1}dx^{p+2}...dx^{n})
\end{equation}%
\begin{equation}
\mathcal{T}\left(  dx^{p+1}...dx^{n}\right)  =i^{\left(  p^{2}\right)
}(dx^{1}dx^{2}...dx^{p})
\end{equation}%
\begin{equation}
\mathcal{T}\left(  dx^{1}...dx^{n}\right)  =1
\end{equation}
we find:%
\begin{align}
\mathcal{T}\left(  dx^{1}...dx^{p}\right)  \bullet\mathcal{T}\left(
dx^{p+1}...dx^{n}\right)   &  =i^{n^{2}}\int_{\mathbb{R}^{0|n}}(\xi
^{p+1}...\xi^{n})(dx^{1}-\xi^{1})...(dx^{p}-\xi^{p})\left[  d^{n}\xi\right]
=\\
i^{n^{2}}\int_{\mathbb{R}^{0|n}}(\xi^{p+1}...\xi^{n})\left(  -1\right)
^{p}\xi^{1}...\xi^{p}\left[  d^{n}\xi\right]   &  =i^{n^{2}}\left(  -1\right)
^{p}\left(  -1\right)  ^{p\left(  n-p\right)  }\mathcal{T}\left(  dx^{1}%
dx^{2}...dx^{n}\right)
\end{align}
\textbf{ }The properties of the convolution reflect on corresponding
properties of the Hodge star operator. Using $\star\omega=i^{\left(
k^{2}-n^{2}\right)  }\mathcal{T}(\omega)$ for $\omega(x,dx)\in\Omega
^{k}(\mathbb{R}^{n})$, we obtain a simple formula for the Hodge dual of the
wedge product of forms in the case $p+q=n$:%
\begin{equation}
\star\left(  \alpha\beta\right)  =\left(  -1\right)  ^{p}\left(  \star
\alpha\right)  \bullet\left(  \star\beta\right)  \label{convoluzionehodge}%
\end{equation}

Considering now the general case of a $p$-form $\alpha$ and a $q$-form $\beta$
in a $n$-dimensional space, one can prove the following relation:
\begin{equation}
\star\left(  \alpha\beta\right)  =\left(  -1\right)  ^{n+q(n-p)}\left(
\star\alpha\right)  \bullet\left(  \star\beta\right)
\label{convoluzionehodgegeneral}%
\end{equation}
\noindent easily checked to be satisfied by the monomials
\begin{equation}
\alpha=dx^{1}dx^{2}...dx^{p},~~~\beta=dx^{n-q+1}dx^{n-q+2}...dx^{n}%
\end{equation}
\noindent Indeed recall that
\begin{equation}
\star\alpha=dx^{p+1}...dx^{n},~~~\star\beta=(-1)^{q(n-q)}dx^{1}...dx^{n-q}%
\end{equation}%
\begin{equation}
\star(\alpha\beta)=(-1)^{q(n-p-q)}dx^{p+1}...dx^{n-q}%
\end{equation}
\noindent Moreover, using the definition of the convolution, one finds
\begin{equation}
(\star\alpha)\bullet(\star\beta)=(-1)^{q(n-q)}(-1)^{p}(-1)^{n(n-q-p)}%
(-1)^{p(n-p)}dx^{p+1}...dx^{n-q}%
\end{equation}
\noindent Comparing the last two equations, relation
(\ref{convoluzionehodgegeneral}) follows. By linearity the same relation
(\ref{convoluzionehodgegeneral}) holds also for generic forms. Two particular
cases provide nontrivial checks:

i) when $\alpha=1\in\Omega^{0}$ :%
\begin{align}
\star\left(  1\beta\right)  =\left(  -1\right)  ^{n+qn}\left(  \star1\right)
\bullet\left(  \star\beta\right)  =\left(  -1\right)  ^{n+qn}\int
_{\mathbb{R}^{0|n}}(\xi^{1}...\xi^{n})\left(  \star\beta(dx-\xi)\right)
\left[  d^{n}\xi\right] \nonumber\\
=\left(  -1\right)  ^{n+qn}\left(  -1\right)  ^{n(n-q)}\star\beta=\star\beta
\end{align}

ii) when $\beta=1\in\Omega^{0}$:%
\begin{align}
\star\left(  \alpha1\right)  =\left(  -1\right)  ^{n}\left(  \star
\alpha\right)  \bullet\left(  \star1\right)  =\left(  -1\right)  ^{n}%
\int_{\mathbb{R}^{0|n}}\left(  \star\alpha\right)  \left(  \xi\right)
(dx^{1}-\xi^{1})...(dx^{n}-\xi^{n})\left[  d^{n}\xi\right] \nonumber\\
=\left(  -1\right)  ^{n}\int_{\mathbb{R}^{0|n}}\left(  \star\alpha\right)
\left(  \xi+dx\right)  \left(  -1\right)  ^{n}\xi^{1}...\xi^{n}\left[
d^{n}\xi\right]  =\star\alpha
\end{align}
where we used the traslational invariance (under $\xi\rightarrow\xi+dx$) of
the Berezin integral.

Similar relations hold (modulo some suitable multiplicative coefficient
depending also on the metric) for the more general integral transform that
gives the Hodge dual for a generic metric $A.$

The convolution defined in the formula (\ref{convoluzione}) can be normalized
as:%
\begin{equation}
\alpha\bullet^{\prime}\beta(x,dx)=(-1)^{\left(  n+pn+pq\right)  }
\alpha\bullet\beta
\end{equation}
where again $p$ is the degree of $\alpha,$ $q$ the degree of $\beta,$ and $n$
the dimension of the space.

With this normalization the formula (\ref{convoluzionehodgegeneral}) looks
better:
\begin{equation}
\star\left(  \alpha\beta\right)  =\left(  \star\alpha\right)  \bullet^{\prime
}\left(  \star\beta\right)  \label{hodgeprodotto}%
\end{equation}
Indeed, noting that $\left(  \star\alpha\right)  \bullet\left(  \star
\beta\right)  =\left(  -1\right)  ^{n+(n-p)n+(n-p)(n-q)}\left(  \star
\alpha\right)  \bullet^{\prime}\left(  \star\beta\right)  $, we have:
\[
\star\left(  \alpha\beta\right)  =\left(  -1\right)  ^{n+q(n-p)}\left(
-1\right)  ^{n+(n-p)n+(n-p)(n-q)}\left(  \star\alpha\right)  \bullet^{\prime
}\left(  \star\beta\right)  =\left(  \star\alpha\right)  \bullet^{\prime
}\left(  \star\beta\right)
\]
The algebra of this new convolution is:
\begin{equation}
\alpha\bullet^{\prime}\beta=(-1)^{\left(  n-p\right)  \left(  n-q\right)
}\beta\bullet^{\prime}\alpha
\end{equation}

\noindent Clearly this normalized convolution product has a unit, the standard
volume form $\star1.$

Equation (\ref{hodgeprodotto}) and the associativity of the wedge product show
that the convolution product $\bullet^{\prime}$ is associative:%
\[
\left(  \star\alpha\bullet^{\prime}\star\beta\right)  \bullet^{\prime}\left(
\star\gamma\right)  =\star\left[  \left(  \alpha\beta\right)  \gamma\right]
=\star\left[  \alpha\left(  \beta\gamma\right)  \right]  =\star\alpha
\bullet^{\prime}\left(  \star\beta\bullet^{\prime}\star\gamma\right)
\]

As last remark, we point out that, using our Fourier representation of the
Hodge dual, it is easy to deduce the standard formula:
\begin{equation}
\alpha\wedge\star\alpha=(\alpha,\alpha)\star1\,,
\end{equation}
where $(\cdot,\cdot)$ is the scalar product associated to the metric $g$.
Moreover, the same scalar product can be rewritten with the new convolution
as
\begin{equation}
(\alpha,\alpha)=\alpha\bullet^{\prime}\star\alpha=(-1)^{p\left(  n-p\right)
}\star\alpha\bullet^{\prime}\alpha
\end{equation}
where instead of the wedge product we have used the convolution product.

\section{Super Hodge dual and super Convolutions}

\noindent We start, as usual, from the real superspace $\mathbb{R}^{n|m}$ with
$n$ bosonic ($x^{i},i=1,\dots,n$) and $m$ fermionic $(\theta^{\alpha}%
,\alpha=1,\dots,m$) coordinates. We denote by $T$ the tangent bundle and by
$T^{\ast}$ the cotangent bundle (see footnote below). To simplify the
notations we will denote by the same letter a bundle and the $\mathbb{Z}_{2}-$graded
modules of its sections.

With ours conventions\footnote{As pointed out in the previous section, in
order to make contact with the standard physical literature we adopt the
conventions that $d$ is an odd operator and $dx$ (an odd form) is dual to the
even vector $\frac{\partial}{\partial x}$. The same holds for the odd
variables $\theta.$ As clearly explained for example in the appendix of the
paper \cite{VORONOV3} if one introduces also the natural concept of even
differential (in order to make more contact with the standard definition of
cotangent bundle of a manifold) our cotangent bundle (that we consider as the bundle 
of one-forms) should, more
appropriately, be denoted by $\Pi T^{\ast}.$} these modules are generated over
the ring of superfunctions as follows $\left(  i=1...n\text{ ; }%
\alpha=1...m\right)  $:
\begin{align*}
&  T\text{ by the even vectors }\frac{\partial}{\partial x^{i}}\ \text{and the
odd vectors }\frac{\partial}{\partial\theta^{\alpha}}\\
&  T^{\ast}\text{ by the even forms }d\theta^{\alpha}\text{and the odd forms
}dx^{i}%
\end{align*}
If $\Pi$ is the parity reversal symbol $\left(  \Pi\mathbb{R}^{p|q}%
=\mathbb{R}^{q|p}\right)  $, we can consider the bundle $\Pi T$. The $\mathbb{Z}_{2}%
-$graded module of its sections is generated by the even vectors $b_{\alpha}$
and the odd vectors $\eta_{i}$

We consider now the $\mathbb{Z}_{2}-$ graded tensor product $T^{\ast}%
\otimes\Pi T\ $and the invariant even section $\sigma$ given by:%
\begin{equation}
\sigma=dx^{i}\otimes\eta_{i}+d\theta^{\alpha}\otimes b_{\alpha}
\label{sezionesigma}%
\end{equation}

Introducing a a (pseudo)riemannian metric $A=g\left(  \frac{\partial}{\partial
x^{i}},\frac{\partial}{\partial x^{j}}\right)  $ and a symplectic form
$B=\gamma(\frac{\partial}{\partial\theta^{\alpha}},\frac{\partial}%
{\partial\theta^{\beta}})$ , the even matrix $\mathbb{G}=%
\begin{pmatrix}
A & 0\\
0 & B
\end{pmatrix}
$ is a supermetric in $\mathbb{R}^{n|m}$ (with obviously $m$ even). $A$ and
$B$ are, respectively, a $n\times n$ matrix and a $m\times m$ matrix with even
entries, $\det A\neq0$ and $\det B\neq0.$

In matrix notations, omitting (here and in the following) the tensor product
symbol, the section $\sigma$ can be written as:%
\[
\sigma=dxAA^{-1}\eta+d\theta BB^{-1}b=dxA\eta^{\prime}+d\theta Bb^{\prime
}=dZ\mathbb{G}W^{\prime}%
\]
where $\eta^{\prime}=A^{-1}\eta$ and $b^{\prime}=B^{-1}b$ are the covariant
forms corresponding to the vectors $\eta$ and $b$, and $dZ=\left(  dx\text{
}d\theta\right)  $ and $W^{\prime}=%
\begin{pmatrix}
\eta^{\prime}\\
b^{\prime}%
\end{pmatrix}
.$

If $\omega(x,\theta,dx,d\theta)$ is a superform in $\Omega^{\left(
p|0\right)  }$ (i.e. a section locally given by a function $\omega
(x,\theta,dx,d\theta)$ with polynomial dependence in the variables $\theta,dx$
and $d\theta,$ of total degree $p$ in the last two variables), the even
section $\sigma$ can be used to generate an integral transform that can be
considered as a fiberwise integration on the fibers of $T^{\ast}$:%

\begin{equation}
\mathcal{T}(\omega)(x,\theta,dx,d\theta)=\int_{\mathbb{R}^{m|n}}%
\omega(x,\theta,\eta^{\prime},b^{\prime})e^{i\left(  dxA\eta^{\prime}+d\theta
Bb^{\prime}\right)  }\left[  d^{n}\eta^{\prime}d^{m}b^{\prime}\right]
\label{trasformatapersuperforme}%
\end{equation}
where $\omega(x,\theta,\eta^{\prime},b^{\prime})$ has polynomial dependence in
the variables $\theta,\eta^{\prime}$ and $b^{\prime}$ and $e^{i\sigma}$ is
defined as the usual power series.

This integral transform clearly depends on the choice of a supermetric and,
from the point of view (relevant for physical applications) of covariance
properties, maps forms to forms. We recall that other important types of
integral transforms (depending on the choice of a volume element but not on a
supermetric) were defined and studied in \cite{Berezin79}, \cite{Voronov-geometric}, \cite{Voronov-quantization}, and  \cite{Voronov-forms}. 

The integral over the odd $\eta^{\prime}$ variables is a Berezin integral and
the integral over the even $b^{\prime}$ variables is defined by formal rules,
for example:%
\begin{subequations}
\begin{align}
\int_{\mathbb{R}^{m}}e^{id\theta Bb^{\prime}}d^{m}b^{\prime}  &  =\frac
{1}{\det B}\,\delta^{m}(d\theta)\label{rappintegrale}\\
\int_{\mathbb{R}^{m}}b_{1}^{\prime}...b_{m}^{\prime}e^{id\theta Bb^{\prime}%
}d^{m}b^{\prime}  &  =(-i)^{m}\,\frac{1}{\left(  \det B\right)  ^{m+1}}\left(
\frac{d}{d\theta}\delta(d\theta)\right)  ^{m} \label{rappintegrale1}%
\end{align}
The products $\delta^{m}(d\theta)$ and $\left(  \frac{d}{d\theta}%
\delta(d\theta)\right)  ^{m}$ ($m$ here denotes the number of factors) are
wedge products ordered as in $d^{m}b.$ In other words this kind of integrals
depends on the choice of an oriented basis. For example, we obtain the crucial
anticommuting property of the delta forms (no sum on $\alpha,\beta$):
\end{subequations}
\begin{equation}
\delta(d\theta^{\alpha})\delta\left(  d\theta^{\beta}\right)  =\int
_{\mathbb{R}^{2}}e^{i(d\theta^{\alpha}b^{\prime\alpha}+d\theta^{\beta
}b^{\prime\beta})}db^{\prime\alpha}db^{\prime\beta}=-\int_{\mathbb{R}^{2}%
}e^{i(d\theta^{\alpha}b^{\prime\alpha}+d\theta^{\beta}b^{\prime\beta}%
)}db^{\prime\beta}db^{\prime\alpha}=-\,\delta(d\theta^{\beta})\delta\left(
d\theta^{\alpha}\right)
\end{equation}

We can generalize the Hodge dual to superforms of zero picture (note that the
spaces of superforms or of integral forms are all finite dimensional) where we
have the two types of differentials, $d\theta$ and $dx.$

A zero picture $p-$superform $\omega$ is a combination of a \textbf{finite
number} of monomial elements of the form:%
\begin{equation}
\rho_{\left(  r,l\right)  }\left(  x,\theta,dx,d\theta\right)  =f(x,\theta
)dx^{i_{1}}dx^{i_{2}}...dx^{i_{r}}\left(  d\theta^{1}\right)  ^{l_{1}}\left(
d\theta^{2}\right)  ^{l_{2}}...\left(  d\theta^{s}\right)  ^{l_{s}}%
\end{equation}
of total degree equal to $p=r+l_{1}+l_{2}+...+l_{s}.$ We denote by $l$ the sum
of the $l_{i}.$ We have also $r\leq n.$

The super Hodge dual on the monomials can be defined as:
\begin{equation}
\star\rho_{\left(  r,l\right)  }=\left(  i\right)  ^{r^{2}-n^{2}}\left(
i\right)  ^{\alpha\left(  l\right)  }\mathcal{T(}\rho_{\left(  r,l\right)
})=\left(  i\right)  ^{r^{2}-n^{2}}\left(  i\right)  ^{\alpha\left(  l\right)
}\frac{\sqrt{\left\vert S\mathrm{\det}\mathbb{G}\right\vert }}{S\mathrm{\det
}\mathbb{G}}\int_{\mathbb{R}^{m|n}}\rho_{\left(  r,l\right)  }(x,\theta
,\eta^{\prime},b^{\prime})e^{i\left(  dxA\eta^{\prime}+d\theta Bb^{\prime
}\right)  }[d^{n}\eta^{\prime}d^{m}b^{\prime}] \label{superhodge1}%
\end{equation}
We recall that:
\[
S\mathrm{\det}\mathbb{G}=\frac{\mathrm{det}A}{\mathrm{det}B}%
\]

The normalization coefficient is given by: $\alpha(l)=2pl-l^{2}-nl-l$ (with
$l=p-r$) if $n$ is even and $\alpha(l)=l$ if $n$ is odd. These coefficient was
computed in \cite{Castellani:2015paa}

The $\star$ operator on monomials can be extended by linearity to generic
forms in $\Omega^{(p|0)}:$%

\[
\star:\Omega^{(p|0)}\longrightarrow\Omega^{(n-p|m)}%
\]
Both spaces are \textbf{finite dimensional} and $\star$ is an
isomorphism\footnote{The normalization coefficients chosen in the definitions
of the duals of $\rho_{\left(  r,l\right)  }$ and $\rho_{\left(  r|j\right)
}$ lead to the usual duality on $\Omega^{\left(  p|0\right)  }:$%
\[
\star\star\rho_{\left(  r,p-r\right)  }=(-1)^{p(p-n)}\rho_{\left(
r,p-r\right)  }%
\]
}.

An important example in $\mathbb{R}^{n|m}$ is $1\in\Omega^{(0|0)}$:%
\[
\star1=\sqrt{\left\vert \frac{\mathrm{det}A}{\mathrm{det}B}\right\vert }%
d^{n}x\delta^{m}(d\theta)\in\Omega^{(n|m)}%
\]

In the case of $\Omega^{(p|m)},$ a $p-$integral\footnote{In the literature, see 
\cite{Bernstein-Leites-Int} and also \cite{Berezin79}, 
one finds pseudodifferential forms of distributional type which 
belong to the spaces $\Omega^{(p|q)}$ where $p$ denote the form degree and $q$ the 
picture number with $0 \leq q \leq m$ (for {\it picture number} we intend the number of Dirac delta functions 
assuming that a given pseudodifferential form can be decomposed in term of them). 
Those with $q=m$ denote the Bernstein-Leites 
integral forms.} form $\omega$ is
a combination of a \textbf{finite number} of monomial elements of the form:%
\begin{equation}
\rho_{\left(  r|j\right)  }\left(  x,\theta,dx,d\theta\right)  =f(x,\theta
)dx^{i_{1}}dx^{i_{2}}...dx^{i_{r}}\delta^{\left(  j_{1}\right)  }\left(
d\theta^{1}\right)  \delta^{\left(  j_{2}\right)  }\left(  d\theta^{2}\right)
...\delta^{\left(  j_{m}\right)  }\left(  d\theta^{m}\right)
\end{equation}
where $p=r-\left(  j_{1}+j_{2}+...+j_{m}\right)  .$ We denote by $j$ the sum
of the $j_{i}.$ We have also $r\leq n$.

The Hodge dual is:%
\begin{equation}
\star\rho_{\left(  r|j\right)  }=\left(  i\right)  ^{r^{2}-n^{2}}\left(
i\right)  ^{\alpha\left(  j\right)  }\frac{\sqrt{\left\vert S\mathrm{\det
}\mathbb{G}\right\vert }}{S\mathrm{\det}\mathbb{G}}\int_{\mathbb{R}^{m|n}}%
\rho_{\left(  r|j\right)  }(x,\theta,\eta,b)e^{i\left(  dxA\eta^{\prime
}+d\theta Bb^{\prime}\right)  }[d^{n}\eta^{\prime}d^{m}b^{\prime}]
\label{superhodge2}%
\end{equation}
Note that we could consider also a more general \textbf{even} super metric:
\begin{equation}
\mathbb{G}=\left(
\begin{array}
[c]{cc}%
G_{(ab)}(x,\theta) & G_{a\beta}(x,\theta)\\
G_{\alpha b}(x,\theta) & G_{[\alpha\beta]}(x,\theta)
\end{array}
\right)  \equiv\left(
\begin{array}
[c]{cc}%
A & C\\
D & B
\end{array}
\right)  \label{superME}%
\end{equation}
where $G_{(ab)}(x,\theta),G_{[\alpha\beta]}(x,\theta)$ are even matrices and
$G_{a\beta}(x,\theta),G_{\alpha b}(x,\theta)$ are odd matrices. In matrix
notation the even section $\sigma$ is in this case given by:%
\[
\sigma=dZ\mathbb{GG}^{-1}W=dxA\eta^{\prime}+d\theta Bb^{\prime}+dxCb^{\prime
}+d\theta D\eta^{\prime}%
\]
In general, the super matrix $\mathbb{G}$ can be expressed in terms of the
supervielbein $\mathbb{V}$ as follows
\[
\mathbb{G}=\mathbb{V}\mathbb{G}_{0}\mathbb{V}^{T}%
\]
where $\mathbb{G}_{0}$ is an invariant constant super matrix characterizing
the tangent space of the supermanifold $\mathbb{R}^{(n|m)}$. The overall
coefficient of the Hodge dual becomes
\[
\frac{\sqrt{|\mathrm{Sdet}\mathbb{G}_{0}|}}{\mathrm{Sdet}\mathbb{V}%
\,\mathrm{Sdet}\mathbb{G}_{0}}%
\]
where $\mathrm{Sdet}\mathbb{V}$ is the superdeterminant of the supervielbein.

In the case of integral forms (that are ``distribution-like") the convolution
product is clearly the right one, because the product of integral forms
vanishes using the graded wedge product. The convolution between two integral
forms $\alpha$ and $\beta$ can be defined as before:%
\begin{equation}
\alpha\bullet\beta(x,dx,\theta,d\theta)=\int_{\mathbb{R}^{n|m}}\alpha
(x,\xi,\theta,b)\beta(x,dx-\xi,\theta,d\theta-b))[d^{n}\xi d^{m}b]
\label{superconvoluzione}%
\end{equation}
Again the integral is a Berezin integral for the variables $\xi$ and a usual
integral for the variables $b.$ In addition, we recall that the definition depends upon a choice of a volume. 
What is important here is that the convolution
of two integral forms is again an integral form (i.e. the total number of
delta forms and derivatives of delta forms is conserved). This convolution
product maps $\Omega^{p|m}\times\Omega^{q|m}\rightarrow\Omega^{p+q-n|m}.$ Note
that, in this case, the form number can be negative.

\section{Integral Representation of PCOs}

Another interesting integral transform (see \cite{VORONOV1}) is a fiberwise
integration not on $T^{\ast}$ but on the parity changed normal bundle $N$ of
the reduced bosonic manifold $M\simeq\mathbb{R}^{n}.$ The reduced manifold $M$
with tangent bundle $T(M)$ is embedded into $\mathcal{M}$ by setting all the
anticommuting coordinates to zero; its normal bundle $N$ is defined as:%
\begin{equation}
0\longrightarrow T(M)\longrightarrow T_{M}\longrightarrow N\longrightarrow0
\end{equation}
where $T_{M}$ is the tangent bundle of the supermanifold $\mathcal{M}$
restricted to the reduced manifold $M.$ The fibers of $N$ are odd and its
sections are generated by the odd vectors $\frac{\partial}{\partial
\theta^{\alpha}}.$ It is known that the total space of $N$ is a supermanifold
isomorphic to $\mathcal{M}$; any real supermanifold can be described as a
vector bundle with even base and odd fibers. We consider now the parity
changed bundle $\Pi N$ with sections generated by the even vectors $b_{\alpha
}$. The total space of $\ \Pi N$ is a bosonic manifold with coordinates $x$
and $b.$ A differential form on $\Pi N$ is a function $\omega(x,b,dx,db)$ and
we can define an integral fiberwise transform:%
\begin{align}
\mathcal{T}(\omega)(x,\theta,dx,d\theta)  &  =\int_{\mathbb{R}^{m|m}}%
\omega(x,b,dx,db)e^{i\left(  \theta db-d\theta b\right)  }\left[
d^{m}(db)d^{m}b\right] \nonumber\\
&  =\int_{\mathbb{R}^{m|m}}\omega(x,b,dx,db)e^{-id\left(  \theta b\right)
}\left[  d^{m}(db)d^{m}b\right]  \label{trasformata2}%
\end{align}
where the even variables $d\theta$ are dual to the even variables $b$ and the
odd variables $\theta$ are dual to the odd variables\ $db.$ The integral over
the odd variables $db$ is a Berezin integral. We see that this integral
transform maps differential forms on the parity changed normal bundle into
pseudoforms on $N$ and does not depend on the choice of a metric.

Depending on the nature of the forms transformed, the pseudoforms obtained can
be very general: polynomial or forms in the commuting differentials and even
distribution-like forms when $\omega(x,b,dx,db)$ is polynomial in the
commuting variables $b.$ If $\omega(x,b,dx,db)$ is generalized to a
distribution valued form its trasformation is a superform\footnote{For
example, if $\omega=bdb$ then $\mathcal{T}(\omega)=i\delta^{\prime}(d\theta)$
and if $\omega=\delta^{\prime}(b)db$ then $\mathcal{T}(\omega)=-id\theta.$ The
imaginary factors could be eliminated introducing a normalization factor in
the definition of the integral transformation.}.

Comparing this transformation with the transformation defined in the formula
(\ref{trasformatapersuperforme}) we observe that it is not suitable for
defining a super Hodge dual but it is connected to the picture changing operators.

The Picture Changing Operators (PCOs) where introduced in \cite{FMS} in string
theory. In supergeometry they were introduced and studied by Belopolsky
(\cite{beloetal1,beloetal2}), For another geometrical interpretation of PCOs as
Poincar\'{e} duals of bosonic submanifolds embedded in a supermanifold see
\cite{Castellani:2014goa} and \cite{SUGRA3D} ). These operators are non
trivial elements of $H_{d}^{(0|r)}$ , act connecting the complexes of (super,
pseudo and integral)-forms on supermanifolds, gives isomorphisms in
cohomology, and can be described in our context as follows.

Given a constant commuting vector $v$ we define the following form:
\begin{equation}
Y_{v}=v_{\alpha}\theta^{\alpha}\delta(v_{\alpha}d\theta^{\alpha})\,,
\label{PCOa}%
\end{equation}
which is $d-$closed but not $d-$exact. $Y_{v}$ belongs to $\Omega^{(0|1)}$ and
by choosing $m$ independent vectors $v^{(\alpha)}$, we have
\begin{equation}
\prod_{\alpha=1}^{m}Y_{v^{(\alpha)}}=\det(v_{\beta}^{(\alpha)})\theta
^{\alpha_{1}}\dots\theta^{\alpha_{m}}\delta(d\theta^{\alpha_{1}})\dots
\delta(d\theta^{\alpha_{m}})\,, \label{PCOb}%
\end{equation}
where $v_{\beta}^{(\alpha)}$ is the $\beta$-component of the $\alpha$-vector.
We can apply the PCO on a given form by taking the graded wedge product.

For example, given $\omega$ in $\Omega^{(p|r)}$ (as discussed in 
footnote 6, $p$ denotes the form degree and $r$ denote the picture number), we have
\begin{equation}
\omega\longrightarrow\omega\wedge Y_{v}\in\Omega^{\left(  {p|r+1}\right)  }\,,
\label{PCOc}%
\end{equation}
Notice that if $r=m$, then $\omega\wedge Y_{v}=0$; on the other hand, if $v$
does not depend on the arguments of the delta forms in $\omega$, we obtain a
non-vanishing form. In addition, if $d\omega=0$ then $d(\omega\wedge Y_{v})=0$
(by applying the Leibniz rule), and if $\omega\neq dK$ then it follows that
also $\omega\wedge Y_{v}\neq dU$ where $U$ is a form in $\Omega^{(p-1|r+1)}$.
The $Y_{v}$ are non trivial elements of the de Rham cohomology and they are
globally defined. So, given an element of the cohomogy $H_{d}^{(p|r)}$, the
new form $\omega\wedge Y_{v}$ is an element of $H_{d}^{(p|r+1)}$.

An integral representation of these operators is obtained acting with the
transformation (\ref{trasformata2}) on suitable forms of type $\omega
(x,dx,b,db).$

For example
\begin{equation}
\mathbb{Y}^{(0|1)}=\theta\delta\left(  d\theta\right)  =\int_{\mathbb{R}%
^{1|1}}e^{i\left(  \theta db-d\theta b\right)  }\left[  d(db)db\right]
\end{equation}
with $\omega(x,dx,b,db)=1$.

The PCO $\mathbb{Y}^{(0|1)}$ is an example of this kind of operators, acting
on $\Omega^{(p|0)}$ and increasing the number of delta forms, and therefore
increasing the picture. As was shown in \cite{Catenacci:2010cs},
$\mathbb{Y}^{(0|1)}$ generates the cohomology $H_{d}^{(0|1)}$ ; therefore, any
other cohomology representative can be expressed in terms of $\mathbb{Y}%
^{(0|1)}$ up to $d$-exact terms. For example, we can consider the following
form
\begin{equation}
\omega(x,dx,b,db)=1+ibdxdb
\end{equation}
whose trasformation gives
\begin{equation}
\widehat{\mathbb{Y}}^{(0|1)}=-dx\delta^{\prime}(d\theta)+\theta\delta
(d\theta)=d(-x\delta^{\prime}(d\theta))+\mathbb{Y}^{(0|1)}\,, \label{pco01}%
\end{equation}
which clearly differs from $\mathbb{Y}^{(0|1)}$ by a $d$-exact term. The new
PCO $\widehat{\mathbb{Y}}^{(0|1)}$ can also be written as
\begin{equation}
\widehat{\mathbb{Y}}^{(0|1)}=-(dx+\theta d\theta)\delta^{\prime}(d\theta).
\end{equation}
It is invariant under the supersymmetry transformations $\delta_{\epsilon
}x=\epsilon\theta$ and $\delta_{\epsilon}\theta=\epsilon$ with $\epsilon$ a
constant anticommuting parameter. Therefore, even though $\widehat{\mathbb{Y}%
}^{(0|1)}$ belongs to the same cohomology class of $\mathbb{Y}^{(0|1)}$, it
has interesting properties, lacking for $\mathbb{Y}^{(0|1)}$. The exact term
in (\ref{pco01}) is not supersymmetric and the corresponding variation of
$\mathbb{Y}^{(0|1)}$ is $d$-exact: $\delta_{\epsilon}\mathbb{Y}^{(0|1)}%
=d(-\epsilon\theta\delta^{\prime}(d\theta))$.

The PCO's of the type $\mathbb{Y}$ are needed to increase the number of
delta's in the differential forms, passing from zero-pictures to the highest
possible picture.\footnote{There is also the possibility to increase the
picture to a number between zero and the maximum value. In that case we have
pseudo-differential forms (i.e. forms with picture $<m$), however, since we do
not use them in the present work, we leave aside such a possibility.} However
given an integral form, we need to be able to construct a superform by acting
with another operator decreasing the picture number. That can be achieved by
considering the following operator:
\begin{equation}
\delta(\iota_{D})=\int_{-\infty}^{\infty}\exp\Big(it\iota_{D}\Big)dt
\end{equation}
where $D$ is an odd vector field on $T(\mathcal{M})$ with $[D,D]\neq0$ and
$\iota_{D}$ is the contraction along the vector $D$.

For example, if we decompose $D$ on a basis $D=D^{\alpha}\partial
_{\theta^{\alpha}}$, where the $D^{\alpha}$ are even coefficients and
$\left\{  \partial_{\theta^{\alpha}}\right\}  $ is a basis of the odd vector
fields, and take $\omega=\omega_{\beta}d\theta^{\beta}\in\Omega^{(1|0)}$, we
have
\begin{equation}
\iota_{D}\omega=D^{\alpha}\omega_{\alpha}=D^{\alpha}\frac{\partial\omega
}{\partial d\theta^{\alpha}}\in\Omega^{(0|0)}\,.
\end{equation}
In addition, due to $[D,D]\neq0$, we have also that $\iota_{D}^{2}\neq0$. The
differential operator $\delta(\iota_{\alpha})\equiv\delta\left(  \iota
_{D}\right)  $ -- with $D=\partial_{\theta^{\alpha}}$ -- acts on the space of
integral forms as follows (we neglect possible introduction of derivatives of
delta forms, but that generalization can be easily done):
\begin{align}
\delta(\iota_{\alpha})\prod_{\beta=1}^{m}\delta(d\theta^{\beta})  &
=\int_{-\infty}^{\infty}\exp\Big(it\iota_{\alpha}\Big)\delta(d\theta^{\alpha
})\prod_{\beta=1\neq\alpha}^{m}\delta(d\theta^{\beta})dt \nonumber\label{exaG}%
\\
&  =\int_{-\infty}^{\infty}\delta(d\theta^{\alpha}+it)\prod_{\beta=1\neq
\alpha}^{m}\delta(d\theta^{\beta})dt=-i\prod_{\beta=1\neq\alpha}^{m}%
\delta(d\theta^{\beta})
\end{align}
The result contains $m-1$ delta forms, and therefore it has picture $m-1$. The
picture number is decreased. Acting several times with $\delta(\iota_{D})$, we
can remove all the delta's. Note that $\delta(\iota_{\alpha})$ is an odd
operator because maps an even (odd) product of delta forms in an odd (even) one.

Let us consider now, in the simplest case $\mathcal{M}=\mathbb{R}^{(1|1)}$,
the following ``double" differential operator
\begin{equation}
\mathbb{Z}^{(0|-1)}=-i\partial_{\theta}\delta(\iota_{_{\partial_{\theta}}%
})=i\delta(\iota_{_{\partial_{\theta}}})\partial_{\theta}%
\end{equation}
where $\partial_{\theta}$ is the partial derivative along $\theta$ and
$\iota_{_{\partial_{\theta}}}$ is the contraction along that vector.

The operator $\mathbb{Z}^{(0|-1)}$ is the product of two operators acting on
different quantities: $\partial_{\theta}$ acts only on functions, and
$\delta(\iota_{_{\partial_{\theta}}})$ acts only on the delta forms. Then, we
can easily check that:
\begin{equation}
\mathbb{Z}^{(0|-1)}\circ\mathbb{Y}^{(0|1)}=1\,. \label{prodZY}%
\end{equation}
A more general form for $\mathbb{Z}^{(0|-1)}$ could be constructed, but we are
not interested here in such a generalization. Moreover, for several variables
$\theta^{\alpha},$ we can consider the product of single operators .

Finally, we note that the Voronov integral transform can be used to produce a
representation of the operator $\mathbb{Z}^{(0|-1)}$ as a multiplication
operator in the space of "dual forms" i.e. of type $\omega(x,b,dx,db)$.

The usual Fourier transform $\mathcal{F}$ in $\mathbb{R}$ (with coordinate
$x)$ gives a representation of the derivative operator $\frac{d}{dx}$ as a
multiplication\footnote{For a function $f(x)$ we have $\left[  \mathcal{F}%
\frac{df}{dx}\right]  (p)=ip\mathcal{F}(f)(p)$ and this is usually written as
$\mathcal{F}\left(  \frac{d}{dx}\right)  =ip$} in the momentum space (with
coordinate $p$).

In our simple case the operator $\mathbb{Z}^{(0|-1)}$ acts on the spaces
$\Omega^{(0|1)}$ and $\Omega^{(1|1)}$ producing elements of $\Omega^{(0|0)}$
and $\Omega^{(1|0)}$ respectively.

A generic form $\omega$ $\in\Omega^{(0|1)}\oplus$ $\Omega^{(1|1)}$ but
$\notin\ker\mathbb{Z}^{(0|-1)}$ can be written as:%
\begin{equation}
\omega(x,dx,\theta,d\theta)=f(x)\theta\delta(d\theta)+g(x)\theta
dx\delta(d\theta).
\end{equation}
because $\mathbb{Z}^{(0|-1)}\Big(\delta(d\theta)\Big)=\mathbb{Z}%
^{(0|-1)}\Big(dx\delta^{\prime}(d\theta)\Big)=\mathbb{Z}^{(0|-1)}\Big(\theta
dx\delta^{\prime}(d\theta)\Big)=0$

The action of the operator $\mathbb{Z}^{(0|-1)}$ gives:%
\begin{equation}
\mathbb{Z}^{(0|-1)}(\omega)=f(x)-g(x)dx\in\Omega^{(0|0)}\oplus\Omega^{(1|0)}%
\end{equation}
Denoting again by $\Pi N$ the even normal bundle of the embedding
$\mathbb{R}\rightarrow\mathbb{R}^{(1|1)}$, acting with the Voronov
\textit{antitransform} (that we call here again $\mathcal{T}$ ) we obtain:%
\begin{equation}
\mathcal{T}(\omega)=f(x)-g(x)dx\in\Omega^{0}\left(  \Pi N\right)  \oplus
\Omega^{1}(\Pi N)
\end{equation}
and
\begin{equation}
\mathcal{T}\Big(\mathbb{Z}^{(0|-1)}(\omega)\Big)(x,dx,b,db)=\left[
f(x)-g(x)dx\right]  \delta\left(  b\right)  db\in\Omega^{0}\left(  \Pi
N\right)  \oplus\Omega^{1}(\Pi N)
\end{equation}
where, as above, we allow also distributional valued differential forms.
Finally, from this we get the desired \textit{right} multiplicative
representation:%
\begin{equation}
\mathcal{T}\Big(\mathbb{Z}^{(0|-1)}\Big)=\delta\left(  b\right)  db\text{ }%
\end{equation}
Similar considerations and computations yield the representation in more
general cases.

\section{Super Liouville Measure for super K\"ahler Manifolds.}

Another interesting integral representation leads to riemannian and symplectic
volumes of supermanifolds. Many examples of riemannian and symplectic volumes
for supermanifolds have been also recently studied by Voronov \cite{VORONOV3}

Let us consider a bosonic compact K\"{a}hler manifold $M$, characterised by a
K\"{a}hler potential $\mathcal{K}$, depending on the complex variables
$(Z^{I})$. From the K\"{a}hler potential, one extracts the K\"{a}hler
$2$-form
\begin{equation}
K^{(2)}=dZ^{I}\wedge d\bar{Z}^{\bar{J}}\partial_{I}\bar{\partial}_{\bar{J}%
}\mathcal{K}=g_{I\bar{J}}(Z,\bar{Z})\,dZ^{I}\wedge d\bar{Z}^{\bar{J}}\,.
\end{equation}
The matrix $g=\left(  g_{I\bar{J}}\right)  $ is a $n\times n$ matrix depending
on $Z^{I}$ and the complex conjugate $\bar{Z}^{\bar{J}}$. The Liouville
measure is given by
\begin{equation}
dV^{(n)}=\underbrace{K^{(2)}\wedge\dots\wedge K^{(2)}}_{n}=\mathrm{det}%
(g)\prod_{I=1}^{n}dZ^{I}d\bar{Z}^{I}\,
\end{equation}
and $\int_{M}dV^{(n)}=\mathrm{vol}(M)$ computes the symplectic volume of the
K\"{a}hler manifold that coincides with the riemannian one.

This expression can be rewritten as a Berezin integral by introducing two sets
of new anticommuting dual variables $\eta^{I},\bar{\eta}^{\bar{I}}$ and
$\xi_{I},\bar{\xi}_{\bar{I}}$
\begin{equation}
\mathrm{det}(g)\prod_{I=1}^{n}dZ^{I}d\bar{Z}^{I}=\int e^{g_{I\bar{J}}%
(Z,\bar{Z})\,\eta^{I}\bar{\eta}^{\bar{J}}+\xi_{I}dZ^{I}+\bar{\xi}_{I}d\bar
{Z}^{I}}\left[  \prod_{I=1}^{n}d\eta^{I}d\bar{\eta}^{\bar{I}}d\xi_{I}d\bar
{\xi}_{\bar{I}}\right]
\end{equation}
The left hand side is a factorized expression and, under a reparametrization
of the manifold, the combination of the two factors is invariant. The right
hand side has the same property as it can be shown by observing that the
variables $\eta^{I}$ transform covariantly, while $\xi_{I}$ transform controvariantly.

Let us move to a supermanifold $\mathcal{M}$ with dimensions $(n|m)$. We
consider a super-K\"{a}hler $2$-superform $K^{(2|0)}$:
\begin{equation}
K^{(2|0)}=g_{I\bar{J}}\,dZ^{I}\wedge d\bar{Z}^{\bar{J}}+g_{I\bar{\beta}%
}\,dZ^{I}\wedge d\bar{\theta}^{\bar{\beta}}+g_{\alpha\bar{J}}\,d\theta
^{\alpha}\wedge d\bar{Z}^{\bar{J}}+g_{\alpha\bar{\beta}}\,d\theta^{\alpha
}\wedge d\bar{\theta}^{\bar{\beta}}%
\end{equation}
where the matrices $g_{I\bar{J}},g_{I\bar{\beta}},g_{\alpha\bar{J}}%
,g_{\alpha\bar{\beta}}$ form a supermatix
\begin{equation}
G_{A\bar{B}}=\left(
\begin{array}
[c]{cc}%
g_{I\bar{J}} & g_{I\bar{\beta}}\\
g_{\alpha\bar{J}} & g_{\alpha\bar{\beta}}%
\end{array}
\right)  \,,~~~~A=(I,\alpha)\,,~~~~\bar{B}=(\bar{J},\bar{\beta})\,,
\end{equation}
whose entries are superfields, functions of $(Z^{I},\theta^{a})$ and of their conjugated.

In this case one cannot simply define the super-Liouville form as
$\underbrace{K^{(2|0)}\wedge\dots\wedge K^{(2|0)}}_{n}$ mainly because this
expression would be a (non integrable) superform of zero picture and not a top
integral form. Using a generalization of the integral representation given
before, one can instead define the correct super-Liouville form as:
\begin{align}
&  \int e^{-(g_{I\bar{J}}\eta^{I}\bar{\eta}^{\bar{J}}+g_{I\bar{\beta}}%
\,\eta^{I}\bar{b}^{\bar{\beta}}+g_{\alpha\bar{J}}\,b^{\alpha}\bar{\eta}%
^{\bar{J}}+g_{\alpha\bar{\beta}}\,b^{\alpha}\bar{b}^{\bar{\beta}})+\xi
_{I}dZ^{I}+\bar{\xi}_{I}d\bar{Z}^{I}+c_{a}d\theta^{\alpha}+\bar{c}%
_{\bar{\alpha}}d\bar{\theta}^{\bar{\alpha}}}\left[  \prod_{I=1}^{n}%
\prod_{\alpha=1}^{m}d\eta^{I}d\bar{\eta}^{\bar{I}}db^{\alpha}d\bar{b}%
^{\bar{\alpha}}d\xi_{I}d\bar{\xi}_{\bar{I}}dc_{\alpha}d\bar{c}_{\bar{\alpha}%
}\right] \label{KaF}\\
&  =\mathrm{Sdet}(G_{A\bar{B}})\prod_{I=1}^{n}dZ^{I}d\bar{Z}^{I}%
\,\prod_{\alpha=1}^{m}\delta(d\theta^{\alpha})\delta(d\bar{\theta}%
^{\bar{\alpha}})\nonumber
\end{align}
That is a top integral form. See also Witten \cite{Witten:2003nn}.

The integral is a Berezin integral in the variables $\eta^{I}$ and $\xi_{I}$
and it is a usual integral over the commuting variables $b^{\alpha}$ and
$c_{\alpha}$. Integrating over $\eta^{I}$ and $\bar{\eta}^{\bar{I}}$ and over
$b^{\alpha}$ and $\bar{b}^{\bar{\alpha}}$ leads to the Berezinian. The
integral over $\xi_{I},\bar{\xi}_{\bar{I}}$ and over $c_{\alpha},\bar{c}%
_{\bar{\alpha}}$ produces the second and the third factor in\ the left hand
side. Both sides are invariant under reparametrization. Notice that the
Berezin integral produces the numerator of the superdeterminant while the
integral over the bosonic variables gives the denominator.
\begin{equation}
\mathrm{Sdet}(G_{A\bar{B}})=\frac{{\det}\left(  g_{I\bar{J}}-g_{I\bar{\beta}%
}(g^{-1})^{\alpha\bar{\beta}}g_{\a\bar{J}}\right)  }{\mathrm{det}\left(
g_{\alpha\bar{\beta}}\right)  }%
\end{equation}

So, the Liouville measure for a K\"{a}hler supermanifold is
\begin{equation}
dV^{(n|m)}=\mathrm{Sdet}(G_{A\bar{B}})\prod_{I=1}^{n}dZ^{I}d\bar{Z}^{I}%
\,\prod_{\alpha=1}^{m}\delta(d\theta^{\alpha})\delta(d\bar{\theta}%
^{\bar{\alpha}})
\end{equation}
The integral of this measure gives the super volume. The volume element
$dV^{(n|m)}$ is an integral top form and (as in the usual bosonic case) it is
equal to the superHodge dual $\star1$ as explained in the previous paragraph.

The same procedure can clearly be implemented in the more general case of a
symplectic supermanifold to produce the \textit{symplectic} volume element,
showing again the power of integral representations for generalizing to
supergeometry many familiar concepts of differential geometry.

\bigskip

\bigskip\bigskip

\textbf{Acknowledgements}

\noindent We thank Paolo Aschieri for valuable discussions.

\end{document}